%% file: main.tex
\documentclass[conference]{IEEEtran}
\IEEEoverridecommandlockouts
\usepackage{cite}
\usepackage{amsmath,amssymb,amsfonts}
\usepackage{algorithmic}
\usepackage{graphicx}
\usepackage{textcomp}
\usepackage{multirow}
\usepackage[table,xcdraw]{xcolor}
\def\BibTeX{{\rm B\kern-.05em{\sc i\kern-.025em b}\kern-.08em
    T\kern-.1667em\lower.7ex\hbox{E}\kern-.125emX}}

\input{macros}

\begin{document}

\title{SkipAnalyzer: A Tool for Static Code Analysis with Large Language Models}
\author{
    \IEEEauthorblockN{
        Mohammad Mahdi Mohajer\IEEEauthorrefmark{1},
        Reem Aleithan\IEEEauthorrefmark{2}, 
        Nima Shiri Harzevili\IEEEauthorrefmark{3},
        Moshi Wei\IEEEauthorrefmark{4},\\
        Alvine Boaye Belle\IEEEauthorrefmark{5},
        Hung Viet Pham\IEEEauthorrefmark{6},
        Song Wang\IEEEauthorrefmark{7}
    }
    \IEEEauthorblockA{Lassonde School of Engineering,
    York University\\
    Toronto, Ontario, Canada\\
    \{\IEEEauthorrefmark{1}mmm98,\IEEEauthorrefmark{2}reem1100,\IEEEauthorrefmark{3}nshiri, \IEEEauthorrefmark{4}moshiwei,\IEEEauthorrefmark{6}hvpham,\IEEEauthorrefmark{7}wangsong\}@yorku.ca;
    \IEEEauthorrefmark{5}alvine.belle@lassonde.yorku.ca
    }
}

\maketitle

\begin{abstract}
\input{sec/abstract}
\end{abstract}


\input{sec/intro}
\input{sec/background}
\input{sec/approach}

\input{sec/experiment}
\input{sec/result}
\input{sec/discussion}
\input{sec/related}

\input{sec/conclusion}

\bibliographystyle{IEEEtran}
\bibliography{paper}

\end{document}

%% file: macros.tex
\usepackage{soul}
\usepackage{hyperref}
\usepackage{pdfpages}
\usepackage{multirow}
\usepackage{listings}
\usepackage{pythonhighlight}
\usepackage{fancybox}
\usepackage{enumitem}
\usepackage{graphicx}
\usepackage{hhline}
\usepackage{color}
\usepackage{xcolor}
\usepackage{array}
\usepackage{amsmath}
\usepackage{xspace}
\usepackage{url}
\usepackage{tikz}
\usepackage{caption}
\usepackage{pgfplots}
\pgfplotsset{compat=1.16}
\usepgfplotslibrary{colorbrewer}
\usepackage{pgf-pie}
\usetikzlibrary{pgfplots.statistics,calc, pgfplots.colorbrewer}
\usepackage{pgfplotstable}
\usepackage{filecontents}

\usepackage{tabularx}
\usepackage[ruled,vlined]{algorithm2e}

\newcommand{\mybox}[1]{%
	\setbox0=\hbox{#1}%
	\setlength{\@tempdima}{\dimexpr\wd0+13pt}%
	\begin{tcolorbox}[boxrule=0.5pt, colback=gray!10, arc=4pt,
		left=6pt,right=6pt,top=6pt,bottom=6pt,boxsep=0pt]
		#1
	\end{tcolorbox}
}

\definecolor{codegreen}{rgb}{0,0.6,0}
\definecolor{codegray}{rgb}{0.5,0.5,0.5}
\definecolor{codepurple}{rgb}{0.58,0,0.82}
\definecolor{backcolour}{rgb}{0.95,0.95,0.92}

\definecolor{dkgreen}{rgb}{0,0.6,0}
\definecolor{gray}{rgb}{0.5,0.5,0.5}
\definecolor{mauve}{rgb}{0.58,0,0.82}
\definecolor{darkblue}{rgb}{0.0,0.0,0.6}
\definecolor{cyan}{rgb}{0.0,0.6,0.6}

\lstdefinestyle{mystyle}{
  language=Java,
  aboveskip=3mm,
  showstringspaces=false,
  columns=flexible,
  numbers=none,
  backgroundcolor=\color{backcolour},
  commentstyle=\color{codegreen},
  keywordstyle=\color{magenta},
  numberstyle=\tiny\color{codegray},
  stringstyle=\color{codepurple},
  basicstyle=\small\ttfamily,
    breakatwhitespace=false,         
    breaklines=false,                 
    captionpos=b,                    
    keepspaces=false,                 
    numbersep=5pt,                  
    showspaces=false,                
    showstringspaces=false,
    showtabs=false,                  
    tabsize=1,
    escapeinside=``
}
\usepackage{tcolorbox}
\makeatletter

\newcommand{\tikzcircle}[2][red,fill=red]{\tikz[baseline=-0.5ex]\draw[#1,radius=#2] (0,0) circle ;}%

\definecolor{songcolor}{RGB}{191,191,191}

\newcommand{\tool}{{SkipAnalyzer}}


%% file: sec/abstract.tex
We introduce {\tool}, a large language model (LLM)-powered tool for static code analysis. It can detect bugs, filter false positive warnings, and patch the detected bugs without human intervention. {\tool} consists of three components, 1) an LLM-based static bug detector that scans source code and reports specific types of bugs, 2) an LLM-based false-positive filter that can identify false-positive bugs {in the results of static bug detectors (e.g., the result of step 1)} to improve detection accuracy, and 3) an LLM-based patch generator that can generate patches for the detected bugs above. As a proof-of-concept, {\tool} is built on ChatGPT, which has exhibited outstanding performance in various software engineering tasks.  

To evaluate {\tool}, we focus on two types of typical and critical bugs that are targeted by static bug detection, i.e., \textit{Null Dereference} and \textit{Resource Leak} as our subjects. We employ Infer to aid the gathering of these two bug types from 10 open-source projects. Consequently, our experiment dataset contains 222 instances of \textit{Null Dereference} bugs and 46 instances of \textit{Resource Leak} bugs.  
Our study demonstrates that {\tool} achieves remarkable performance in the mentioned static analysis tasks, including bug detection, false-positive warning removal, and bug repair. 
In static bug detection, {\tool} achieves accuracy values of up to 68.37\% for detecting \textit{Null Dereference} bugs and 76.95\% for detecting \textit{Resource Leak} bugs, improving the precision of the current leading bug detector, Infer by 12.86\% and 43.13\% respectively.
For removing false-positive warnings, {\tool} can reach a precision of up to 93.88\% for \textit{Null Dereference} bugs and 63.33\% for \textit{Resource Leak} bugs. 
Additionally, {\tool} surpasses state-of-the-art false-positive warning removal tools. 
Furthermore, in bug repair, {\tool} can generate syntactically correct patches to fix its detected bugs with a success rate of up to 97.30\%. 

%% file: sec/intro.tex
\section{Introduction}
\label{sec:intro}

Numerous static code analysis techniques have been utilized in the literature for the automatic detection of real-world software bugs~\cite{phoenix-apr-sa-violation,static-apr, avatar-sa-violation, sa-violation-sonarqube, c3pr}. These tools typically rely on predefined heuristic rules to scan and analyze the codebases or binaries of software projects \cite{sa-violation-sonarqube, c3pr, avatar-sa-violation}. During this analysis, any violations of these rules are categorized as a bug, leading the tools to flag the corresponding code artifact, such as a line or a group of lines, as buggy. However, employing static bug detectors  presents specific challenges. One primary issue is that most static bug detectors generate numerous false-positive warnings \cite{survey-static-vuldetect-tse, emp-effectiveness-sa-c-vuldetect}. Consequently, additional manual review is essential to validate the reported potential bugs, resulting in a time-consuming and labor-intensive process \cite{survey-android-malware-detection-sa}. 
{Additionally, these detected bugs still require manual fixes, demanding developers' expertise and knowledge, which can be a resource-intensive task \cite{apr-era-llm, apr-survey}.}
{Recently, Large Language Models (LLMs) such as ChatGPT have demonstrated significant potential in various reasoning and decision-making roles, serving as intelligent agents~\cite{wang2023voyager,song2023llm}, especially in SE tasks such as code generation, understanding, and debugging. However, to date, there has yet to be research exploring the capabilities of LLM-based tools for various static code analysis tasks. 
To address this gap, in this paper, we take a step toward developing a tool for code analysis based on LLMs named {\tool}. It contains three components: 1) an LLM-based static bug detector that is capable of scanning source code and reporting specific types of bugs, 2) an LLM-based false-positive filter for identifying false-positive bugs {(e.g., those from step 1)} to improve the detection accuracy, and 3) an LLM-based patch generator that can generate patches for the detected bugs. As a proof-of-concept, {\tool} is built on ChatGPT, which has consistently demonstrated outstanding performance in various software engineering tasks~\cite{llm4se-survey}.  

To evaluate {\tool}, we select two typical and widely-studied bugs commonly targeted by static bug detection: \textit{Null Dereference} and \textit{Resource Leak} as our subjects. 
{Following existing works\cite{kharkar2022learning, harzevili2022characterizing}}, we utilize Infer to facilitate the collection of these two types of bugs from 10 open-source projects. As a result, our experiment dataset includes 222 \textit{Null Dereference} bugs and 46 \textit{Resource Leak} bugs.  
When employing ChatGPT, we leverage two different model versions: 1) ChatGPT-3.5 Turbo and 2) ChatGPT-4. 
In this study, we adhere to the best practices of prompt engineering \cite{chatgpt-promptdesign-dl-repair, prompting-is-all-you-need} and create precise and context-aware prompts to effectively harness the language models' capabilities. We explore various prompting strategies, including zero-shot, one-shot, and few-shot prompting, to evaluate the performance of different ChatGPT models across various scenarios. By strategically adapting our prompts and methodologies, we aim to identify the most efficient and accurate ways of leveraging LLMs in static analysis.


{Our experiments reveal that {\tool} achieves remarkable performance with significant improvements when compared to previous baseline methods for each respective task: 1) Its LLM-based static bug detector achieves a precision rate that is 12.86\% higher for \textit{Null Dereference} bugs and 43.13\% higher for \textit{Resource Leak} bugs compared to Infer. 2) Its LLM-based false-positive filter can enhance Infer's precision rate by 28.68\% for \textit{Null Dereference} bugs and 9.53\% for \textit{Resource Leak} bugs, surpassing the performance of existing baselines. {Additionally, it can enhance the precision of {\tool}'s first component by 16.31\% for \textit{Null Dereference} bugs.} 3) Its LLM-based patch generator can generate patches for the buggy codes and repair them with correctness rates of 97.30\% and 91.77\% for \textit{Null Dereference} and \textit{Resource Leak} bugs, respectively. Furthermore, on average, 98.77\% of the generated code for both \textit{Null Dereference} and \textit{Resource Leak} bugs are syntactically correct.}


As a summary, this paper contributes to the field in the following ways:

\begin{itemize}
    \item We propose {\tool}, a large language
    model (LLM) powered tool for static code analysis that can detect bugs, filter false positive warnings, and patch the detected bugs without human intervention. 

    \item We show that {\tool} can be an effective static bug detector, improving the base results of the current state-of-the-art tool, Infer.
    
    \item We demonstrate that {\tool} can effectively eliminate false-positive warnings from the output of static bug detectors such as Infer and {its first component}, thereby enhancing their precision and surpassing the performance of existing baseline methods in false-positive warning removal.
    
    \item We illustrate that {\tool} exhibits the capability to repair bugs identified by static bug detectors through the generation of accurate patches, which we compare against a baseline approach from prior research.


    \item We release the dataset and the source code for our experiments for future usages\footnote{\url{https://doi.org/10.5281/zenodo.10043170}}.
    
\end{itemize}

The structure of the remainder of this paper is as follows: 
Section \ref{sec:bg} presents the background of the study.
Section \ref{sec:data} presents the data collection of the study.
Section \ref{sec:setup} describes the proposed approach. 
Section \ref{sec:experiment} presents the experiment settings. 
Section \ref{sec:result} presents the results of our work. 
Section \ref{sec:discussion} presents the discussion. 
Section \ref{sec:related} presents the related work.
Section \ref{sec:threats} presents the threats to validity, and Section \ref{sec:conclusion} concludes the paper.

%% file: sec/background.tex
\section{Background}
\label{sec:bg}

\subsection{Static Bug Detection}
\label{sec:sa}

Static bug detection is an automated technique for inspecting and analyzing a program's source code, object code, or binaries, all without executing the program \cite{sa-android-survey, comparative-sa-java}. This process identifies potential bugs by examining how the code's control and data flow align with specific bug patterns and rules \cite{dev-engange-sa, comparative-sa-java}. If a code section violates any of these rules, the static bug detector will issue a warning concerning the violated rule for that particular piece of code \cite{sa-npe-ase}. 
Multiple tools and methods have been created to perform static bug detection within the body of research and industry \cite{harzevili2023}. Infer, created by Meta, is a static bug detection tool capable of being utilized across various programming languages, including Java, C, C++, Objective-C, and C\#. It accomplishes this by utilizing a predetermined set of rules to identify potential bugs and conducting inter-procedural analysis as part of the project compilation process\cite{InferRSS}. SpotBugs \cite{spotbugs-wild}, an enhanced iteration of findBugs \cite{findbugs, findbugs2}, employs a methodology similar to Infer. It leverages the concept of bug patterns, which consist of specific rules and templates designed to identify particular types of bugs. That being said, applying SpotBugs is limited only to Java byte codes and does not support other languages' source codes or binaries. Google has also introduced ErrorProne, a static bug detector tailored for Java programs \cite{error-prone}. ErrorProne is designed to catch common programming errors and potential bugs during compilation. It enhances the compiler's type analysis, enabling developers to identify more mistakes before advancing to production. {In this study, we employ Infer as a baseline for comparison with \tool's first component and as the tool for generating warnings for our data collection.}


\subsection{False-Positive Warning Removal}
\label{sec:fpremoval}

A significant issue associated with the use of static bug detectors is their tendency to generate a considerable volume of inaccurate warnings, which are essentially alerts that are not genuine indicators of actual bugs \cite{kharkar2022learning, detecting-false-alarms-davidlo, ml-based-fpwremoval2, ml-based-fpwremoval3, efindbugs-error-ranking, identify-fp-parttern-sa, techniques-eff-automated-elim-fp, slr-actionable-alert}. Recent research demonstrates that the false-positive warning rate can escalate to as high as 91\% \cite{detecting-false-alarms-davidlo}.
Also, removing high false positive warnings in static analysis tools is very time-consuming for developers since it requires them to verify the generated warnings manually, and it often results in frustration and diminished utilization of these tools \cite{what-dev-want-program-analysis, why-not-sa}. Therefore, exploring strategies for reducing false positive warnings in static bug detectors is crucial to increasing the developer's trust and confidence while using such tools \cite{why-not-sa, emp-ml-triage-reports-java-sa}. 
Recent studies have addressed this issue by providing various techniques in detecting and eliminating false-positive warnings. Junker et al. \cite{junker2012smt} introduced a method to address false-positive warnings by transforming this issue into a syntactic model-checking problem and employing SMT solvers to evaluate the feasibility of violation of formula in model checking as a counter-example. Wang et al. \cite{golden-feature} have proposed a ``Golden Features" set to detect actionable warnings and eliminate the unactionable ones. Hanam et al. \cite{ml-based-fpwremoval2} proposed an approach based on machine learning, where they created a warning prediction model to distinguish between actionable and non-actionable warnings. 
Recently, Kharkar et al.~\cite{kharkar2022learning} introduced distinct tools that leverage state-of-the-art neural models, mostly transformer-based models, which are widely regarded as the most effective approach for eliminating false-positive warnings. {In this work, we opt to utilize the tools outlined in this study as our baselines for comparison. These tools are referred to as feature-based approach, DeepInferEnhance, and a GPT-C powered approach \cite{kharkar2022learning}.}


\subsection{Automated Program Repair}
\label{sec:repair}
Automated program repair (APR) techniques have recently garnered significant attention from researchers \cite{automation-fixing-sb, chatgpt-promptdesign-dl-repair, apr-era-llm, codex-apr-improving}. The core concept of program repair is to automatically generate program fixes to facilitate the testing and validation of software systems \cite{apr-survey, automatic-finding-patches-gen-programming}. The APR tools usually take two inputs, a flawed code and a localization of the bug in the flawed code or a description of the program's expected behaviour \cite{dynamic-apr-formal-methods}. Automated program repair can be both done dynamically and statically. In the dynamic setting, the localization or description of the bug usually comes in the test suites associated with that flawed code and should be executed to test the validation of the generated patch by the tool \cite{dynamic-apr-formal-methods, apr-era-llm}. As an example, Liu et al. \cite{phoenix-apr-sa-violation} introduced a method known as Phoenix, which utilizes fix patterns derived from the static analysis of bug violations to generate patches. Their research involved the utilization of the Defects4J benchmark dataset, which includes test suites used to validate the effectiveness of the generated patches for the faulty code.
Furthermore, Xia et al. \cite{apr-era-llm} investigated utilizing Large Language Models (LLMs) for the first time for Automated Program Repair (APR). They examined nine recent and advanced LLMs using five benchmark datasets. All the patches they generated for the faulty codes were validated against the accompanying test suites in their respective datasets.
In a static context, the bug's description or localization is not provided through test suites; instead, it is presented using alternative heuristic or formal methods \cite{phoenix-apr-sa-violation, static-apr, vulrepair}. For example, Tonder and Goues \cite{static-apr} proposed an APR technique for fixing general pointer safety properties using Separation Logic \cite{sep-logic}. Bavishi et al. \cite{phoenix-apr-sa-violation} introduced a method for generating patches and verifying them using a static analyzer as the oracle, eliminating the need for a test suite. Fu et al. \cite{vulrepair} introduced VulRepair, a neural Automated Program Repair (APR) method founded on the CodeT5 model \cite{codet5}. Their research involved generating patches for localized bugs found in a pre-existing dataset \cite{bigvul, cvefixes}. They evaluated the effectiveness of their approach using a perfect prediction metric and compared the results to the ground truth data within the same dataset. 

{In this study, we select VulRepair \cite{vulrepair} as our baseline tool to compare with \tool. This choice is made because VulRepair, similar to our approach, does not necessitate the execution of test cases for validation.}






\subsection{Large Language Models}
\label{sec:llm}

Large Language Models (LLMs) have gained significant popularity in recent research and industrial applications. Numerous recent studies are investigating the utilization of LLMs in the field of Software Engineering (SE), driven by the significant progress and advancements achieved by LLMs \cite{codex-apr-improving, apr-era-llm, extensive-pretrained-program}. ChatGPT \cite{chatgpt}, one of the most renowned LLMs, has gained widespread recognition recently in performing software engineering tasks\cite{prompting-is-all-you-need,apr-era-llm, chatgpt-promptdesign-dl-repair, chatgpt-automated-code-refinement}. ChatGPT is accessible throughout an API\footnote{\url{https://platform.openai.com/docs/api-reference}} and has been created by harnessing the capabilities of two state-of-the-art GPT models, specifically, GPT-3.5 Turbo \cite{gpt-3.5} and GPT-4 \cite{gpt-4}. Utilizing LLMs like ChatGPT as decision making components introduces a novel approach to systematically interact with instruction-tuned LLMs, a method known as prompt engineering. Prompt engineering is the practice of creating tailored input queries that effectively communicate with LLMs\cite{prompting-is-all-you-need, chatgpt-promptdesign-dl-repair}. Numerous investigations leverage prompt engineering in their utilization of LLMs \cite{llm-fewshot-tester, llm-nl2code-survey, llm-survey, chatgpt-promptdesign-dl-repair, apr-era-llm}. Prompt engineering as a practice offers the flexibility to utilize various strategies, including the zero-shot approach, where the LLM is prompted without any prior input/output examples; the one-shot method, involving an additional example; and the few-shot strategy, denoted as K-shot, which provides K examples as previous input/output pairs for the LLM \cite{llm-fewshot-tester, prompting-is-all-you-need}. Moreover, in prompt engineering, techniques like Chain-of-Thought (COT) are employed to enhance the correctness of generated output. This is achieved by either including the thinking steps in examples or requesting an explanation of the decision-making process from the LLM \cite{cot, prompting-is-all-you-need, zs-cot}.

%% file: sec/approach.tex
\section{Data Collection}
\label{sec:data}
In this work, 
we take two types of typical and critical bugs that are targeted by static bug detection, i.e., \textit{Null Dereference} and \textit{Resource Leak} as our subjects. To accelerate the data collection, we first applied Infer \cite{InferRSS} to our experimental projects with a focus on detecting \textit{Null Dereference} and \textit{Resource Leak} bugs.
The rationale behind selecting Infer is its extensive adoption in various companies, including Microsoft. Furthermore, Infer exhibits higher precision in comparison to alternative static analysis tools, leading to the generation of more valid warnings~\cite{kharkar2022learning}. 
We apply Infer to a selection of seven prominent GitHub projects (with at least more than 200 stars), along with three projects featured in prior research, to generate warnings \cite{kharkar2022learning}. These projects are shown in Table \ref{table:projects}.

\input{table/projects}

Note that, as Infer can report false positives~\cite{harzevili2022characterizing, kharkar2022learning}, for each warning, reported, we further manually check whether it is a true bug and not a false positive. 
This manual labeling process involves three authors with at least four years of development experience, each independently reviewing all the reported warnings by Infer. For each warning, they assign a binary label, i.e., zero indicating a ``false positive'', signifying that the warning generated by Infer is incorrect and does not represent a true bug, and one indicating a ``true positive'', indicating that the warning is accurate and demonstrates a real bug. 
Following this individual labeling, the authors then collaborate to identify and resolve any discrepancies or conflicts in their assessments. After resolving these conflicts, we have our comprehensive dataset containing warnings generated by Infer, along with their corresponding ground truth labels. 
{Also, for each of the true bugs, the participants work together to create a patch that can fix the bug.}

\section{Approach}
\label{sec:setup}


Figure \ref{fig:overview} provides an overview of {\tool}'s pipeline, which contains 1) an LLM-based static bug
detector that can scan source code and report specific types of bugs (Section~\ref{sec:step1}), 
2) an LLM-based false positive filter that can identify false-positive bugs in the result of step 1 for improving the detection accuracy  (Section~\ref{sec:step2}), 
and 3) an LLM-based patch generator that can generate patches for the detected bugs  (Section~\ref{sec:step3}).


\subsection{LLM-based Static Bug Detector}
\label{sec:step1}

In the first component, {\tool} takes the buggy code snippet as input for in-depth analysis by the LLM. The LLM has been given a specialized prompt with specifications for each bug type to increase its detection validity. For example, to address \textit{Null Dereference} bugs, we collect common bug patterns for this type of bug, like not having null checks before dereferencing an object. We then provide this information to the LLM in the initial prompt. This specialization helps the LLM understand the task at hand. Additionally, for Resource Leak bugs, a similar approach is used. Moreover, specific structured output requirements are defined to facilitate easy parsing and extracting necessary information from the LLM's responses.
Moreover, the prompt of this component supports adding additional examples to apply one-shot or few-shot strategies.
Given that we possess a ground truth for the provided buggy code snippet, we expect {\tool} to recognize the problem previously identified by Infer and issue a valid warning for it. Furthermore, {\tool} offers an additional explanation for each potential bug it detects and the warnings it generates.

\subsection{LLM-based False-Positive Warning Filter}
\label{sec:step2}

In the second component, {\tool}'s objective is to improve the precision of static bug detection by eliminating false-positive warnings. To achieve this, {\tool} takes both the buggy code snippet and the warning associated with that code snippet, which is generated by a static bug detector. These warnings can originate from various static bug detectors, such as Infer or {\tool} itself. Subsequently, the LLM, previously instructed with specialized guidelines, evaluates the buggy code snippet and its corresponding warning. Similar to the first component, the prompt in this stage allow for the inclusion of additional examples, enabling the application of one-shot or few-shot strategies. 

\subsection{LLM-based Static Bug Repair}
\label{sec:step3}

In the third component, {\tool} processes the buggy code snippet and the warning related to that particular bug in the code. It then proceeds to generate a patch for the buggy code, ultimately producing the fixed version of the code. This component feeds the inputs mentioned above to the LLM, which has previously received tailored instructions on addressing our specific types of bugs. It is worth noting that the prompt of this component does not support additional examples for one-shot or few-shot strategies since examples in our dataset usually have multiple implementations of methods with varying lengths and violate the maximum token limitation of our opted LLMs (more details in Section \ref{sec:subjects}).

\input{figure/overview}

%% file: table/projects.tex
\begin{table*}[t]
\caption{Summary of analyzed projects. In this table, projects highlighted in \tikzcircle[yellow, fill=yellow]{2pt} are from a recent study done by Kharkar et al. \cite{kharkar2022learning}, and projects highlighted in \tikzcircle[orange, fill=orange]{2pt} are the ones we collected from the popular repositories. The warnings reported in this table are generated by Infer \cite{InferRSS} and manually verified.
}
\label{table:projects}
\centering
\renewcommand{\arraystretch}{1.4}
\begin{tabular}{cc|c|c|c|c|c|}
\cline{2-7}
\multicolumn{1}{c|}{} &
  \textbf{Project} &
  \textbf{Project Description} &
  \textbf{Version} &
  \textbf{LOC} &
  \textbf{Repository Group} &
  \textbf{Number of Verified Warnings} \\ \cline{2-7} 
\multicolumn{1}{c|}{} &
  \cellcolor[HTML]{FFFFC7}nacos &
  Dynamic naming and configuration service &
  2.0.2 &
  217,653 &
  Alibaba &
  58 \\ \cline{2-7} 
\multicolumn{1}{c|}{} &
  \cellcolor[HTML]{FFFFC7}azure-maven-plugins &
  \begin{tabular}[c]{@{}c@{}}Maven plugins for Azure services\end{tabular} &
  2.2.2 &
  53,025 &
  Microsoft &
  45 \\ \cline{2-7} 
\multicolumn{1}{c|}{\multirow{-3}{*}{}} &
  \cellcolor[HTML]{FFFFC7}playwright-java &
  \begin{tabular}[c]{@{}c@{}}Java library to automate Chromium, \\ Firefox, and WebKit with a single API\end{tabular} &
  1.13.0 &
  67,548 &
  Microsoft &
  5 \\ \cline{2-7} 
\multicolumn{1}{c|}{} &
  \cellcolor[HTML]{FFCE93}java-debug &
  \begin{tabular}[c]{@{}c@{}}Java Debug Server, an implementation of \\ Visual Studio Code (VSCode) Debug Protocol\end{tabular} &
  0.47.0 &
  22,852 &
  Microsoft &
  2 \\ \cline{2-7} 
\multicolumn{1}{c|}{} &
  \cellcolor[HTML]{FFCE93}dolphinscheduler &
  Modern data orchestration platform &
  2.0.9 &
  215,808 &
  Apache &
  100 \\ \cline{2-7} 
\multicolumn{1}{c|}{} &
  \cellcolor[HTML]{FFCE93}dubbo &
  \begin{tabular}[c]{@{}c@{}}high-performance, Java-based \\ open-source RPC framework\end{tabular} &
  3.2 &
  350,957 &
  Apache &
  193 \\ \cline{2-7} 
\multicolumn{1}{c|}{} &
  \cellcolor[HTML]{FFCE93}bundletool &
  \begin{tabular}[c]{@{}c@{}}a tool to manipulate Android App Bundles\\  and Android SDK Bundles\end{tabular} &
  1.15.1 &
  135,711 &
  Google &
  51 \\ \cline{2-7} 
\multicolumn{1}{c|}{} &
  \cellcolor[HTML]{FFCE93}guava &
  \begin{tabular}[c]{@{}c@{}}set of core Java libraries from Google \\ that includes new data structures\end{tabular} &
  32.1.1 &
  698,201 &
  Google &
  35 \\ \cline{2-7} 
\multicolumn{1}{c|}{} &
  \cellcolor[HTML]{FFCE93}jreleaser &
  \begin{tabular}[c]{@{}c@{}}release automation tool for \\ Java and non-Java projects\end{tabular} &
  1.7.0 &
  114,914 &
  Community &
  30 \\ \cline{2-7} 
\multicolumn{1}{c|}{\multirow{-7}{*}{}} &
  \cellcolor[HTML]{FFCE93}jsoup &
  \begin{tabular}[c]{@{}c@{}}Java library for working with \\ HTML Document Object Model (DOM)\end{tabular} &
  1.16.1 &
  33,689 &
  Community &
  33 \\ \cline{2-7} 
 &
  \multicolumn{2}{c|}{} &
  \multicolumn{3}{c|}{\cellcolor[HTML]{EFEFEF}\textbf{Total Number of Verified Warnings}} &
  \cellcolor[HTML]{EFEFEF}\textbf{552} \\ \cline{4-7} 
\end{tabular}

\end{table*}

%% file: figure/overview.tex
\begin{figure}[t!]
\begin{center}
    
\includegraphics[width=0.45\textwidth]{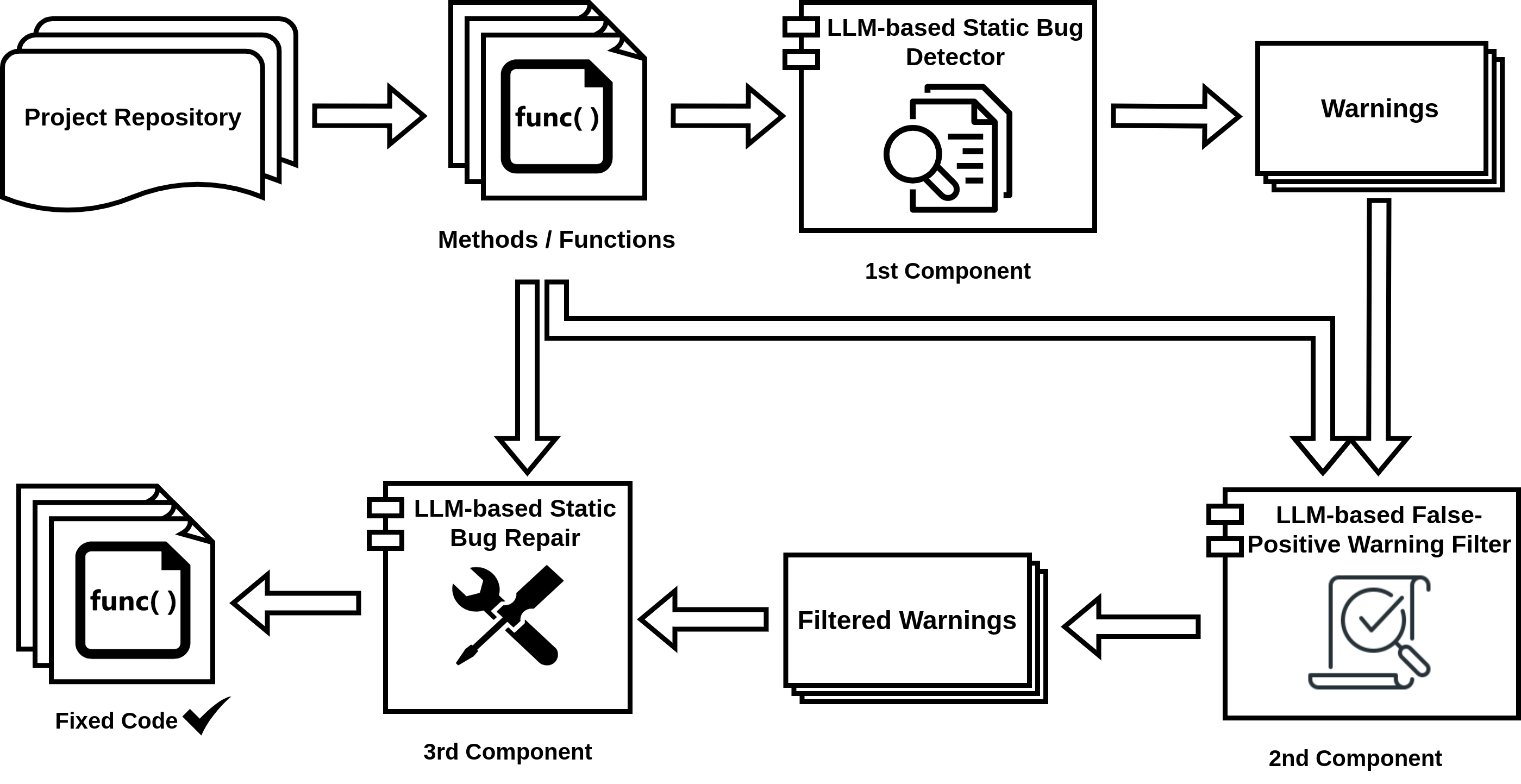}
    
\end{center}

\caption{The overview of \tool.} 
\label{fig:overview}
\end{figure}

%% file: sec/experiment.tex
\section{Experiment Settings}
\label{sec:experiment}


\subsection{Research Questions}
\label{sec:rqs}

To evaluate the performance of {\tool}, we design experiments to answer the following research questions:

\vspace{4pt}
\noindent \textbf{RQ1 (Static Bug Detection): What is the performance of {\tool} in detecting bugs?} 
In this research question, we explore the potential of {\tool} in the realm of static bug detection. We present the performance of our approach for static bug detection and compare it with our baseline, Infer, while also conducting comparisons under various prompting strategies.

\vspace{4pt}
\noindent \textbf{RQ2 (False Positive Warning Removal): What is the effectiveness of {\tool} in filtering false positive warnings?}
In this research question, we delve into the effectiveness of {\tool}'s second component in identifying and eliminating false-positive warnings from the output of a static bug detector. This analysis is applied to both the warnings generated by Infer and those produced in response to RQ1. Eventually, we compare the results of our developed approach with recent state-of-the-art baselines provided by Kharkar et al.~\cite{kharkar2022learning} that utilize feature-based and neural models for false-positive warning removal.

\vspace{4pt}
\noindent \textbf{RQ3 (Static Bug Repair): What is {\tool}'s performance in repairing a buggy code for a specific type of bug?}
This RQ examines the potential of {\tool} in repairing instances of bugs identified. 
Note that, as we do not have test cases for the identified bugs, most generate-and-validate program repair techniques cannot be our baselines, as these approaches require a set of test cases as the ground truth. 
In this work, we choose VulRepair~\cite{vulrepair}, a CodeT5-based static automated program repair model~\cite{wang2021codet5} that does need test cases as the ground truth.

\subsection{ChatGPT Versions}
\label{sec:subjects}
We choose the latest variations of ChatGPT models \cite{chatgpt}, ChatGPT-4 \cite{gpt-4} and ChatGPT-3.5 Turbo \cite{gpt-3.5}, as the primary backbone to build {\tool}. These models have demonstrated strong performance in various software engineering tasks, including code generation, code comprehension, and debugging \cite{apr-era-llm, llm-survey}. We interact with these models using OpenAI API\footnote{\href{https://platform.openai.com/docs/introduction}{https://platform.openai.com/docs/introduction}}. According to the model documentation, there is a wide range of options available for each of the models, and each option comes with its distinct characteristics, including varying token lengths, associated costs, and update frequencies \cite{chatgpt}. For ChatGPT-4 and ChatGPT-3.5 Turbo, we utilize the default settings, which provide maximum token support of 8,192 and 4,097, respectively \cite{gpt-3.5, gpt-4}. 
Due to this limitation in the maximum input token, we constrain the code inputs to the method scope, implying that we only provide the methods themselves as input code snippets to the LLMs. 

\subsection{Prompting Strategies}
\label{sec:prompting}

In each of the {\tool}'s components, we use different prompt engineering strategies such as zero-shot, one-shot, and few-shot strategies. In the zero-shot strategy, the LLM is prompted without preceding examples from the previous LLM input and output pairs. In contrast, the few-shot and one-shot strategies incorporate examples from the previous LLM input and output pairs. The difference between the one-shot and few-shot strategies lies in the number of examples included: the one-shot strategy employs a single example in the prompt, while the few-shot strategy encompasses multiple examples. 
In the first component (LLM-based static bug detector) and the second component (LLM-based false-positive warning filter), we use zero-shot, one-shot, and few-shot strategies. 
In this paper, for the few-shot strategy, (\textit{K}-shot), we input the model with three examples (\textit{K} = 3). The rationale for selecting \textit{K} = 3 is based on the consideration that opting for values exceeding three could potentially violate the maximum token limit constraint imposed on our inputs for ChatGPT models (details in Section \ref{sec:subjects}). 
In the third component (LLM-based bug repair), we only use the zero-shot strategy due to the limitations mentioned in Sections \ref{sec:subjects} and \ref{sec:step3}.
Furthermore, in the prompts for all of {\tool}'s components, we request the LLM to explain its decision-making process and the steps it takes to arrive at its conclusions. According to the literature, this approach, known as zero-shot Chain-of-Thought reasoning, can enhance the output's robustness and validity \cite{cot, zs-cot}.

\subsection{Data Sampling}



Our experiments employ \textit{N}-fold cross-validation to remove potential data sampling bias, with \textit{N} set to 5 in this study~\cite{kfold}. This approach involves splitting the dataset into five subsets, where one-fifth of the data serves as the validation set, and the remaining four-fifths function are used for selecting examples in the process of prompting the LLMs for one-shot and few-shot strategies. We select the examples for these strategies randomly in a uniform distribution. 
Also, in each of the examples utilized in one-shot and few-shot strategies, we incorporate one instance of a true-positive record and one instance of an false-positive record to prevent any bias towards a particular group of examples when applying these strategies to the LLM.



\input{table/RQ1}

\subsection{Evaluation Metrics}

We employ the following evaluation metrics to assess our experiments concerning each of our research questions: 

For \textbf{Static Bug Detection (RQ1)} and \textbf{False-Positive Warning Removal (RQ2)}, since we have a ground truth for the warnings we generated for our dataset for evaluation, we use Accuracy, Precision, and Recall metrics.
Also, to choose the best combination of strategy and model, we use the F1-score metric. 
For \textbf{Static Bug Repair (RQ3)}, we assess the performance using two key measures that we devised: ``Logic Rate'', which signifies the proportion of repaired code with correct logic, and ``Syntax Rate'', indicating the proportion of repaired code with correct syntax. To evaluate the ``Logic Rate'', we assess correctness by comparing the repaired code to manually crafted ground truth fix patches prepared by our team. In the case of the ``Syntax Rate'', we employ a Java parser to validate the syntax of the generated code.

%% file: table/RQ1.tex
\begin{table*}[h]
\caption{Summary of {\tool}'s results for each of the model-strategy combination in Static Bug Detection (RQ1) using two datasets. In this table, rows highlighted in \tikzcircle[yellow, fill=yellow]{2pt} indicate the most effective combination of model and strategy for \textit{Null Dereference} bugs, and rows in \tikzcircle[cyan, fill=cyan]{2pt} indicate the most effective one for \textit{Resource Leak} bugs.
}
\label{table:rq1}
\centering

\begin{tabular}{|c|c|c|c|c|c|c|c|c|}
\hline
 &
   &
   &
   &
   &
   &
   &
   &
   \\
\multirow{-2}{*}{\textbf{Dataset}} &
  \multirow{-2}{*}{\textbf{Bug Type}} &
  \multirow{-2}{*}{\textbf{Infer's Precision}} &
  \multirow{-2}{*}{\textbf{Strategy}} &
  \multirow{-2}{*}{\textbf{Model}} &
  \multirow{-2}{*}{\textbf{Accuracy}} &
  \multirow{-2}{*}{\textbf{Precision}} &
  \multirow{-2}{*}{\textbf{Recall}} &
  \multirow{-2}{*}{\textbf{F1-Score}} \\ \hline
 &
   &
   &
  \cellcolor[HTML]{FFFFC7} &
  GPT-3.5-Turbo &
  50.66\% &
  52.32\% &
  40.64\% &
  45.75\% \\ \cline{5-9} 
 &
   &
   &
  \multirow{-2}{*}{\cellcolor[HTML]{FFFFC7}\textbf{Zero Shot}} &
  \cellcolor[HTML]{FFFFC7}\textbf{GPT-4} &
  \cellcolor[HTML]{FFFFC7}\textbf{68.37\%} &
  \cellcolor[HTML]{FFFFC7}\textbf{63.76\%} &
  \cellcolor[HTML]{FFFFC7}\textbf{88.93\%} &
  \cellcolor[HTML]{FFFFC7}\textbf{74.27\%} \\ \cline{4-9} 
 &
   &
   &
   &
  GPT-3.5-Turbo &
  60.67\% &
  62.49\% &
  59.76\% &
  61.09\% \\ \cline{5-9} 
 &
   &
   &
  \multirow{-2}{*}{One Shot} &
  GPT-4 &
  64.54\% &
  62.20\% &
  79.18\% &
  69.67\% \\ \cline{4-9} 
 &
   &
   &
   &
  GPT-3.5-Turbo &
  60.47\% &
  66.14\% &
  48.63\% &
  56.05\% \\ \cline{5-9} 
 &
  \multirow{-6}{*}{Null Dereference} &
  \multirow{-6}{*}{50.9\%} &
  \multirow{-2}{*}{Few Shot} &
  GPT-4 &
  64.31\% &
  65.21\% &
  64.96\% &
  65.09\% \\ \cline{2-9} 
 &
   &
   &
  \cellcolor[HTML]{96FFFB} &
  GPT-3.5-Turbo &
  56.31\% &
  44.34\% &
  40.44\% &
  42.30\% \\ \cline{5-9} 
 &
   &
   &
  \multirow{-2}{*}{\cellcolor[HTML]{96FFFB}\textbf{Zero Shot}} &
  \cellcolor[HTML]{96FFFB}\textbf{GPT-4} &
  \cellcolor[HTML]{96FFFB}\textbf{76.95\%} &
  \cellcolor[HTML]{96FFFB}\textbf{82.73\%} &
  \cellcolor[HTML]{96FFFB}\textbf{55.11\%} &
  \cellcolor[HTML]{96FFFB}\textbf{66.15\%} \\ \cline{4-9} 
 &
   &
   &
   &
  GPT-3.5-Turbo &
  34.87\% &
  35.03\% &
  72.22\% &
  47.18\% \\ \cline{5-9} 
 &
   &
   &
  \multirow{-2}{*}{One Shot} &
  GPT-4 &
  72.78\% &
  68.39\% &
  63.55\% &
  65.88\% \\ \cline{4-9} 
 &
   &
   &
   &
  GPT-3.5-Turbo &
  49.31\% &
  41.49\% &
  65.55\% &
  50.82\% \\ \cline{5-9} 
\multirow{-12}{*}{Our collection} &
  \multirow{-6}{*}{Resource Leak} &
  \multirow{-6}{*}{39.6\%} &
  \multirow{-2}{*}{Few Shot} &
  GPT-4 &
  75.25\% &
  74.12\% &
  59.55\% &
  66.04\% \\ \hhline{=========}
 &
   &
   &
  \cellcolor[HTML]{FFFFC7} &
  GPT-3.5-Turbo &
  50.31\% &
  69.85\% &
  41.53\% &
  52.09\% \\ \cline{5-9} 
 &
   &
   &
  \multirow{-2}{*}{\cellcolor[HTML]{FFFFC7}\textbf{Zero Shot}} &
  \cellcolor[HTML]{FFFFC7}\textbf{GPT-4} &
  \cellcolor[HTML]{FFFFC7}\textbf{77.90\%} &
  \cellcolor[HTML]{FFFFC7}\textbf{81.87\%} &
  \cellcolor[HTML]{FFFFC7}\textbf{85.51\%} &
  \cellcolor[HTML]{FFFFC7}\textbf{83.65\%} \\ \cline{4-9} 
 &
   &
   &
   &
  GPT-3.5-Turbo &
  58.25\% &
  70.90\% &
  59.87\% &
  64.92\% \\ \cline{5-9} 
 &
   &
   &
  \multirow{-2}{*}{One Shot} &
  GPT-4 &
  70.51\% &
  80.95\% &
  72.56\% &
  76.53\% \\ \cline{4-9} 
 &
   &
   &
   &
  GPT-3.5-Turbo &
  56.47\% &
  75.60\% &
  51.28\% &
  61.11\% \\ \cline{5-9} 
 &
  \multirow{-6}{*}{Null Dereference} &
  \multirow{-6}{*}{65.2\%} &
  \multirow{-2}{*}{Few Shot} &
  GPT-4 &
  68.52\% &
  80.57\% &
  67.56\% &
  73.50\% \\ \cline{2-9} 
 &
   &
   &
  \cellcolor[HTML]{96FFFB} &
  GPT-3.5-Turbo &
  61.66\% &
  50.0\% &
  50.0\% &
  50\% \\ \cline{5-9} 
 &
   &
   &
  \multirow{-2}{*}{\cellcolor[HTML]{96FFFB}\textbf{Zero Shot}} &
  \cellcolor[HTML]{96FFFB}\textbf{GPT-4} &
  \cellcolor[HTML]{96FFFB}\textbf{84.61\%} &
  \cellcolor[HTML]{96FFFB}\textbf{100\%} &
  \cellcolor[HTML]{96FFFB}\textbf{71.42\%} &
  \cellcolor[HTML]{96FFFB}\textbf{83.32\%} \\ \cline{4-9} 
 &
   &
   &
   &
  GPT-3.5-Turbo &
  46.66\% &
  50\% &
  70\% &
  58.33\% \\ \cline{5-9} 
 &
   &
   &
  \multirow{-2}{*}{One Shot} &
  GPT-4 &
  73.33\% &
  60\% &
  50\% &
  54.54\% \\ \cline{4-9} 
 &
   &
   &
   &
  GPT-3.5-Turbo &
  48.33\% &
  36.66\% &
  40\% &
  38.26\% \\ \cline{5-9} 
\multirow{-12}{*}{Projects from \cite{kharkar2022learning}} &
  \multirow{-6}{*}{Resource Leak} &
  \multirow{-6}{*}{53.8\%} &
  \multirow{-2}{*}{Few Shot} &
  GPT-4 &
  68.33\% &
  60\% &
  40\% &
  48\% \\ \hline
\end{tabular}
\end{table*}

%% file: sec/result.tex
\section{Result Analysis}
\label{sec:result}

This section presents the results and answers the research questions we asked in Section \ref{sec:rqs}.

\subsection{RQ1: Performance of {\tool} on Static Bug Detection}
\label{sec:rq1}



\noindent \textbf{Approach:} In the case of static bug detection, we input the code snippet of each record in our dataset to the {\tool}'s first component. We expect that {\tool} will identify the issue previously described and detected by Infer and produce a valid warning. We perform our experiments under different prompting strategies such as zero-shot, one-shot, and few-shot strategies (more details in Section \ref{sec:prompting}) by using two different ChatGPT models, i.e., ChatGPT-3.5 Turbo and ChatGPT-4 (more details in Section \ref{sec:subjects}).

\noindent \textbf{Baselines:} As a baseline for {\tool}'s first component, we select Infer, a state-of-the-art static bug detector. Using our collected dataset, we compare {\tool}'s performance in static bug detection with Infer. 



\noindent \textbf{Result:} Table \ref{table:rq1} summarizes the results 
of {\tool}'s static bug detection task {under different prompting strategies with different ChatGPT models} using Infer on our collected dataset and the projects used in the prior study by Kharkar et al~\cite{kharkar2022learning}. 
We also show the optimal combination of model and strategy for the static bug detection task for each of the datasets, which is the one that has the highest F1-Score for a specific bug type. Notably, in our collected dataset, the most effective combination for both \textit{Null Dereference} and \textit{Resource Leak} bugs involves utilizing ChatGPT-4 in conjunction with the zero-shot strategy. Considering this, in static bug detection, {\tool} can achieve accuracy, precision, and recall rates of 68.37\%, 63.76\%, and 88.93\%, respectively, for \textit{Null Dereference} bugs. Likewise, for \textit{Resource Leak bugs}, these metrics can attain values of 76.95\%, 82.73\%, and 55.11\%, respectively. This indicates that {\tool} exhibits precision levels that are 12.86\% and 43.13\% higher than Infer, surpassing the performance of the state-of-the-art static bug detector baseline. Furthermore, it is noteworthy that some other model-strategy combinations also demonstrate superior performance when compared to Infer. For instance, ChatGPT-3.5-Turbo with the one-shot strategy continues to outperform Infer in the detection of \textit{Null Dereference} bugs, achieving a rate of 62.49\% compared to Infer's 50.9\%.
We can also see a similar result in the dataset from the projects used by Kharkar et al. \cite{kharkar2022learning}. In this scenario, the most effective strategy and model combination is the zero-shot strategy coupled with the ChatGPT-4 model. Compared to Infer's base results on this dataset depicted in Table \ref{table:rq1}, we have a 16.6\% and 46.2\% boost in precision for \textit{Null Dereference} and \textit{Resource Leak} bugs, respectively.






\mybox{\textbf{Answer to RQ1:} SkipAnalyzer can significantly enhance the state-of-the-art static bug detection tool (i.e., Infer) on detecting \textit{Null Dereference} and \textit{Resource Leak} bugs.} bugs.

\subsection{RQ2: Performance of {\tool} on False-Positive Warning Removal}
\label{sec:rq2}

\noindent \textbf{Approach:} For answering this RQ, we input both a code snippet and the warning linked to the code snippet generated by a bug detector to {\tool}. 
To examine the generalizability of {\tool} in removing false positives, we use the warnings generated by both Infer and {\tool} (as described in Section \ref{sec:rq1}). We also perform our experiments under different prompting strategies such as zero-shot, one-shot, and few-shot strategies (more details in Section \ref{sec:subjects}) by using two different ChatGPT models, i.e., ChatGPT-3.5 Turbo and ChatGPT-4 (more details in Section \ref{sec:subjects}).


\noindent \textbf{Baselines:} We select the state-of-the-art false-positive removal approach proposed in the recent study conducted by Kharkar et al.~\cite{kharkar2022learning}, i.e., GPT-C and also they created two baselines to evaluate GPT-C, which are a feature-based logistic regression model~\cite{kharkar2022learning} and DeepInferEnhance (based on CodeBERTa)~\cite{kharkar2022learning}. 
However, the authors neither disclosed the source code for their tool nor their experiments due to confidential issues. They also did not release their collected dataset. Thus, to enable a meaningful comparison with the baseline tools outlined in their study, we gathered data that closely mirrored the one described in their paper, including the same projects and similar versions. Ultimately, we conducted our experiments on a dataset akin to the one they utilized, allowing for a fair and direct comparison between our tool and their baseline tools.

\noindent \textbf{Result:} As we explained, we have two options for the static bug detector utilized for false-positive warning removal:

\subsubsection{Option 1 -- Infer}

Table \ref{table:rq2-infer-warnings} shows the results of applying SkipAnalyzer as a false-positive warning removal tool on warnings generated by Infer and {\tool}.

\input{table/RQ2-infer}

As demonstrated, when dealing with \textit{Null Dereference} bugs, we can enhance Infer's precision by 28.68\% by employing the zero-shot strategy alongside the ChatGPT-4 model. In the case of \textit{Resource Leak} bugs, Infer's precision can be improved by up to 9.53\% when utilizing the zero-shot strategy combined with the ChatGPT-3.5-Turbo model. Furthermore, in comparison to the current baselines~\cite{kharkar2022learning}, as indicated in Table \ref{table:rq2-baselines}, our findings reveal that {\tool} can surpass the existing baselines in terms of precision improvement, with a margin of at least 11.21\% and 3.97\% for \textit{Null Dereference} and \textit{Resource Leak} bugs, respectively. Also, it is worth noting that Kharkar et al. \cite{kharkar2022learning} did not provide the results of the feature-based logistic regression model and DeepInferEnhance for \textit{Resource Leak} bugs. Therefore, we specified them with ``--'' in Table \ref{table:rq2-baselines}.


\input{table/RQ2-baselines}

\subsubsection{Option 2 -- SkipAnalyzer}

Table \ref{table:rq2-infer-warnings} also shows the result of false-positive warning removal on the warnings generated by {\tool} in Section \ref{sec:rq1}. Given that these warnings may arise from various model-strategy combinations, we selected the most effective model-strategy combination for warning generation further to enhance precision through the false-positive warning removal process. Therefore, we opted for the zero-shot strategy with the ChatGPT-4 model.

Moreover, it is worth noting that no false-positive warnings were associated with \textit{Resource Leak} issues in the warnings generated by this specific model-strategy combination. Consequently, our primary focus remains improving the false-positive warnings related to \textit{Null Dereference} issues.

As we can see in Table \ref{table:rq2-infer-warnings}, by selecting a proper model-strategy combination, which is, in this case, the few-shot strategy with the ChatGPT-4 model, we can improve the precision by removing false-positive warnings of the warnings generated in Section \ref{sec:rq1} by 16.31\%. 


\mybox{\textbf{Answer to RQ2:} {\tool} can remove false-positive warnings effectively and it outperforms the previous state-of-the-art false-positive warning removal baselines.}

\subsection{RQ3: Performance of {\tool} on Bug Repair}
\label{sec:rq3}

\noindent \textbf{Approach:} 
For this RQ, we input {\tool} with the true positive buggy code snippets. We instruct the {\tool} to understand the code and its corresponding warning and generate a potential patch to repair the buggy code. Subsequently, we manually assess the generated code patches for their syntactical and logical correctness. We perform our experiments under the zero-shot strategy by using two different ChatGPT models, i.e., ChatGPT-3.5 Turbo and ChatGPT-4 (more details in Section \ref{sec:subjects}).
It is important to note that we do not employ the one-shot or few-shot strategy for static bug repair due to the limitations mentioned in Sections \ref{sec:step3} and \ref{sec:subjects}.

\noindent \textbf{Baselines:} We compare the results of our tool with a recent baseline tool called VulRepair \cite{vulrepair}. We used the model and experiment codes offered by the authors to test this baseline tool on our collected dataset. We followed the same configuration for training and evaluating their model on our dataset.

\noindent \textbf{Result:} Our research findings indicate that {\tool} can serve as a robust bug repair tool. Table \ref{table:rq3} displays the performance of SkipAnalyzer in fixing \textit{Null Dereference} and \textit{Resource Leakage} bugs for different models. The results reveal that {\tool} surpasses the baseline tool, VulRepair \cite{vulrepair}, with a substantial improvement. Specifically, {\tool} achieves a logic rate increase of up to 78.91\% for \textit{Null Dereference} bugs and a rate increase of 78.87\% for Resource Leak bugs compared to VulRepair. 
Furthermore, it is worth noting that {\tool} does not require training or fine-tuning. In contrast, VulRepair mandates the adaptation and training of our dataset for patch generation, showing the superiority of {\tool} over the recent state-of-the-art baseline tool, VulRepair. Also, VulRepair-generated patches are not self-contained Java codes and should be appended to the buggy code. Hence, they are not parsable, and we specify the Syntax Rate for VulRepair's generated patches with \ref{sec:rq3} in Table \ref{table:rq3}.

\input{table/RQ3}


\mybox{\textbf{Answer to RQ3:} {\tool} is effective in patching the detected bugs. In addition, {\tool} can significantly outperform a recent state-of-the-art baseline tool, i.e., VulRepair.}

%% file: table/RQ2-infer.tex
\begin{table*}[t!]
\caption{Summary of {\tool}'s results in False-Positive Warning Removal (RQ2) on warnings generated by Infer and {\tool}. In this table, \textbf{$\text{P}_{\text{Original}}$} is the precision of the static bug detector for the corresponding bug type and \textbf{$\text{P}_{\text{After}}$} is the precision of the static bug detector after applying False-Positive Warning Removal process. ``Imp.'' indicates the amount of precision improvement. 
Rows highlighted in \tikzcircle[yellow, fill=yellow]{2pt} indicate the most effective combination of model and strategy for \textit{Null Dereference} bugs, and rows in \tikzcircle[cyan, fill=cyan]{2pt} indicate the most effective one for \textit{Resource Leak} bugs. Also, the records that have no improvement in precision are shown with ``--''.
} 
\label{table:rq2-infer-warnings}
\centering
\begin{tabular}{|c|c|c|c|c|c|c|c|c|c|}
\hline
\textbf{Static Bug Detector} &
  \textbf{Bug Type} &
  \textbf{Strategy} &
  \textbf{Model} &
  \textbf{$\text{P}_{\text{Original}}$} &
  \textbf{$\text{P}_{\text{After}}$} &
  \textbf{Imp.} &
  \textbf{Recall} &
  \textbf{Accuracy} &
  \textbf{F1-Score} \\ \hline
 &
   &
   &
  GPT-3.5-Turbo &
  \cellcolor[HTML]{FFFFC7} &
  40\% &
  - &
  13.07\% &
  37.91\% &
  19.71\% \\ \cline{4-4} \cline{6-10} 
 &
   &
  \multirow{-2}{*}{Zero Shot} &
  GPT-4 &
  \cellcolor[HTML]{FFFFC7} &
  95\% &
  \textbf{+29.8} &
  27.30\% &
  51.41\% &
  42.42\% \\ \cline{3-4} \cline{6-10} 
 &
   &
   &
  GPT-3.5-Turbo &
  \cellcolor[HTML]{FFFFC7} &
  59.32\% &
  - &
  43.71\% &
  43.91\% &
  50.33\% \\ \cline{4-4} \cline{6-10} 
 &
   &
  \multirow{-2}{*}{One Shot} &
  GPT-4 &
  \cellcolor[HTML]{FFFFC7} &
  95\% &
  \textbf{+29.8} &
  51.66\% &
  66.29\% &
  66.93\% \\ \cline{3-4} \cline{6-10} 
 &
   &
  \cellcolor[HTML]{FFFFC7} &
  GPT-3.5-Turbo &
  \cellcolor[HTML]{FFFFC7} &
  53.63\% &
  - &
  52.17\% &
  41.40\% &
  52.89\% \\ \cline{4-4} \cline{6-10} 
 &
  \multirow{-6}{*}{Null Dereference} &
  \multirow{-2}{*}{\cellcolor[HTML]{FFFFC7}\textbf{Few Shot}} &
  \cellcolor[HTML]{FFFFC7}\textbf{GPT-4} &
  \multirow{-6}{*}{\cellcolor[HTML]{FFFFC7}65.2\%} &
  \cellcolor[HTML]{FFFFC7}\textbf{93.88\%} &
  \cellcolor[HTML]{FFFFC7}\textbf{+28.68} &
  \cellcolor[HTML]{FFFFC7}\textbf{64.23\%} &
  \cellcolor[HTML]{FFFFC7}\textbf{73.46\%} &
  \cellcolor[HTML]{FFFFC7}\textbf{76.27\%} \\ \cline{2-10} 
 &
   &
  \cellcolor[HTML]{96FFFB} &
  \cellcolor[HTML]{96FFFB}\textbf{GPT-3.5-Turbo} &
  \cellcolor[HTML]{96FFFB} &
  \cellcolor[HTML]{96FFFB}\textbf{63.33\%} &
  \cellcolor[HTML]{96FFFB}\textbf{+9.53} &
  \cellcolor[HTML]{96FFFB}\textbf{80\%} &
  \cellcolor[HTML]{96FFFB}\textbf{75\%} &
  \cellcolor[HTML]{96FFFB}\textbf{70.69\%} \\ \cline{4-4} \cline{6-10} 
 &
   &
  \multirow{-2}{*}{\cellcolor[HTML]{96FFFB}\textbf{Zero Shot}} &
  GPT-4 &
  \cellcolor[HTML]{96FFFB} &
  60\% &
  \textbf{+6.2\%} &
  60\% &
  80\% &
  60\% \\ \cline{3-4} \cline{6-10} 
 &
   &
   &
  GPT-3.5-Turbo &
  \cellcolor[HTML]{96FFFB} &
  40\% &
  - &
  60\% &
  50\% &
  48\% \\ \cline{4-4} \cline{6-10} 
 &
   &
  \multirow{-2}{*}{One Shot} &
  GPT-4 &
  \cellcolor[HTML]{96FFFB} &
  60\% &
  \textbf{+6.2\%} &
  50\% &
  75\% &
  54.54\% \\ \cline{3-4} \cline{6-10} 
 &
   &
   &
  GPT-3.5-Turbo &
  \cellcolor[HTML]{96FFFB} &
  50\% &
  - &
  80\% &
  50\% &
  61.53\% \\ \cline{4-4} \cline{6-10} 
\multirow{-12}{*}{Infer} &
  \multirow{-6}{*}{Resource Leak} &
  \multirow{-2}{*}{Few Shot} &
  GPT-4 &
  \multirow{-6}{*}{\cellcolor[HTML]{96FFFB}53.8\%} &
  60\% &
  \textbf{+6.2\%} &
  60\% &
  80\% &
  60\% \\ \hhline{==========}
 &
   &
   &
  GPT-3.5-Turbo &
  \cellcolor[HTML]{FFFFC7} &
  70\% &
  - &
  19.27\% &
  32.93\% &
  30.22\% \\ \cline{4-4} \cline{6-10} 
 &
   &
  \multirow{-2}{*}{Zero Shot} &
  GPT-4 &
  \cellcolor[HTML]{FFFFC7} &
  86\% &
  \textbf{+4.13\%} &
  26.72\% &
  37.45\% &
  40.78\% \\ \cline{3-4} \cline{6-10} 
 &
   &
   &
  GPT-3.5-Turbo &
  \cellcolor[HTML]{FFFFC7} &
  67\% &
  - &
  30.72\% &
  32.72\% &
  42.26\% \\ \cline{4-4} \cline{6-10} 
 &
   &
  \multirow{-2}{*}{One Shot} &
  GPT-4 &
  \cellcolor[HTML]{FFFFC7} &
  97.5\% &
  \textbf{+15.63} &
  52.72\% &
  59.92\% &
  68.44\% \\ \cline{3-4} \cline{6-10} 
 &
   &
  \cellcolor[HTML]{FFFFC7} &
  GPT-3.5-Turbo &
  \cellcolor[HTML]{FFFFC7} &
  73.97\% &
  - &
  38.18\% &
  38.90\% &
  50.36\% \\ \cline{4-4} \cline{6-10} 
\multirow{-6}{*}{\begin{tabular}[c]{@{}c@{}}SkipAnalyzer \\ \end{tabular}} &
  \multirow{-6}{*}{Null Dereference} &
  \multirow{-2}{*}{\cellcolor[HTML]{FFFFC7}\textbf{Few Shot}} &
  \cellcolor[HTML]{FFFFC7}\textbf{GPT-4} &
  \multirow{-6}{*}{\cellcolor[HTML]{FFFFC7}81.87\%} &
  \cellcolor[HTML]{FFFFC7}\textbf{98.18\%} &
  \cellcolor[HTML]{FFFFC7}\textbf{+16.31} &
  \cellcolor[HTML]{FFFFC7}\textbf{77.45\%} &
  \cellcolor[HTML]{FFFFC7}\textbf{80.25\%} &
  \cellcolor[HTML]{FFFFC7}\textbf{86.59\%} \\ \hline
\end{tabular}
\end{table*}

%% file: table/RQ2-baselines.tex
\renewcommand{\arraystretch}{1.4}
\begin{table}[h]
\caption{Performance of False-Positive Warning Removal of baseline tools proposed by \cite{kharkar2022learning}. In this table column ``Precision Imp.'' indicates the precision improvement of Infer after applying each of the baseline tools. Also, column ``S.A. Recall Imp.'' shows {\tool} 's improvement in Recall for each bug type compared to the baseline tool. 
}
\label{table:rq2-baselines}
\centering
\resizebox{0.5\textwidth}{!}{
\begin{tabular}{|c|c|c|c|c|}
\hline
\textbf{Baseline}                 & \textbf{Bug Type} & \textbf{Precision Imp.} & \textbf{Recall} & \textbf{SA. Recall Imp.} \\ \hline
\multirow{2}{*}{Feature-based}    & Null Dereference  & +8.26\%                 & 65.1\%          & +12.35\%                 \\ \cline{2-5} 
                                  & Resource Leak     & --                      & --              & --                       \\ \hline
\multirow{2}{*}{DeepInferEnhance} & Null Dereference  & +15.13\%                & 88.3\%          & -10.85\%                 \\ \cline{2-5} 
                                  & Resource Leak     & --                      & --              & --                       \\ \hline
\multirow{2}{*}{GPT-C}            & Null Dereference  & +17.47\%                & 83.7\%          & -6.25\%                    \\ \cline{2-5} 
                                  & Resource Leak     & +5.56\%                 & 64.5\%          & +15.5\%                  \\ \hline
\end{tabular}
}
\end{table}

%% file: table/RQ3.tex
\begin{table}[t!]
\caption{Summary of SkipAnalyzer's results in static bug repair. The ``Syntax Rate'' column indicates the proportion of generated patches that are syntactically correct. ``Logic Rate'' column represents the rate of generated patches that are logically correct.} 
\label{table:rq3}
\centering
\resizebox{0.5\textwidth}{!}{
\begin{tabular}{|c|c|c|c|}
\hline
\textbf{Bug Type}                  & \textbf{Model}                         & \textbf{Logic Rate}                      & \textbf{Syntax Rate}                     \\ \hline
                                   & GPT-3.5-Turbo                          & 94.25\%                                  & 100.0\%                                  \\ \cline{2-4} 
                                   & \cellcolor[HTML]{FFFFC7}\textbf{GPT-4} & \cellcolor[HTML]{FFFFC7}\textbf{97.30\%} & \cellcolor[HTML]{FFFFC7}\textbf{99.55\%} \\ \cline{2-4} 
\multirow{-3}{*}{Null Dereference} & VulRepair                              & 18.39\%                                  & --                                        \\ \hline
                                   & GPT-3.5-Turbo                          & 87.11\%                                  & 97.77\%                                  \\ \cline{2-4} 
                                   & \cellcolor[HTML]{96FFFB}\textbf{GPT-4} & \cellcolor[HTML]{96FFFB}\textbf{91.77\%} & \cellcolor[HTML]{96FFFB}\textbf{97.77\%} \\ \cline{2-4} 
\multirow{-3}{*}{Resource Leak}    & VulRepair                              & 12.90\%                                  & --                                        \\ \hline
\end{tabular}
}
\end{table}

%% file: sec/discussion.tex
\section{Discussions}
\label{sec:discussion}
\subsection{Reasons for Missing Detecting Bugs}

{\tool} also has limitations such as it can miss detecting bugs in RQ1. In this section, we explore the underlying reasons behind the bugs that {\tool} cannot detect. 
For our analysis, we first collect all the missing bugs that {\tool} can not detect. Then, we manually examine all the missing bugs and observe the possible recurring and common patterns among them. 
\input{figure/discussion/ND1}

\input{figure/discussion/ND3}

\subsubsection{Null Dereference}
We have identified three patterns from \textit{Null Dereference} bugs that {\tool} missed. Firstly, {\tool} struggles to distinguish \textit{Null Dereferences} of objects within a null check for another object. This is often observed when {\tool} overlooks objects that might become null and dereferenced later within a null check for another object. This issue becomes more pronounced when these objects have some form of relationship with one another, such as when the first object serves as an argument to a method call of the second object. Figure~\ref{fig:discussion-nd1} shows an example bug that is missing by {\tool}. 

Second, we have observed that {\tool} sometimes makes random assumptions about the method, API call, class variable, and instance variable that are usually outside the current method's scope. {\tool}'s inability to access the implementation details of these entities hinders its ability to determine whether they can return null or not. Consequently, {\tool} occasionally makes incorrect assumptions about the return values of such entities, leading to missing potential \textit{Null Dereference} bugs. Figure \ref{fig:discussion-nd3} shows an example bug that cannot be detected by {\tool} because of the above reason.

Third, another prevalent pattern is {\tool}'s difficulty in detecting \textit{Null Dereferences} that occur in a single line through a chain of method invocations. While {\tool} may be able to detect the first object invoking a method, it often fails to identify the subsequent methods chained through method calls to the initial object, which could also return null. As an example, we can look at Figure \ref{fig:discussion-nd4}.

\input{figure/discussion/ND4}

\subsubsection{Resource Leak}

The prevalent pattern of overlooked \textit{Resource Leak} bugs involves cases where the code partially handles the release of resources. This implies that while some resources are properly released after their use, there are still other resources that remain unreleased. Figure \ref{fig:discussion-rl1} is an example of a missing \textit{Resource Leak} bug due to this issue.

\input{figure/discussion/RL1}



%% file: figure/discussion/ND1.tex
\begin{figure}[t!]
\begin{center}
    
\begin{lstlisting}[
    numberstyle=\bfseries\small, % Make line numbers bold and small
    numbersep=0pt,   % Adjust the separation between code and line numbers
    numbers=left,
    basicstyle=\small, %or \small or \footnotesize etc.
]
public DependentParameters getDependency() {
   if (this.dependency == null) {
     Map<String, Object> taskParamsMap = 
     JSONUtils.parseObject(this.getTaskParams(), new 
     TypeReference<Map<String, Object>>() {});
     this.dependency = JSONUtils.parseObject((String) 
     taskParamsMap.get(Constants.DEPENDENCE), 
     DependentParameters.class);
  }
    return this.dependency;
}
\end{lstlisting}

\fbox{
    \begin{minipage}{8.5cm}
        \textbf{Ground Truth Warning:} object ``taskParamsMap'' last assigned on line 3 could be null and is dereferenced at line 7. 
    \end{minipage}
}

\fbox{
    \begin{minipage}{8.5cm}
        \textbf{{\tool}'s Explanation:} The code checks if ``this.dependency'' is null before assigning a value to it. Therefore, there is no potential null dereference bug in this code.
    \end{minipage}
}

\end{center}

\caption{An example of a missing \textit{Null Dereference}.}
\label{fig:discussion-nd1}
\end{figure}

%% file: figure/discussion/ND3.tex
\begin{figure}[t!]
\begin{center}
    
\begin{lstlisting}[
    numberstyle=\bfseries\small, % Make line numbers bold and small
    numbersep=0pt,   % Adjust the separation between code and line numbers
    numbers=left,
    basicstyle=\small, %or \small or \footnotesize etc.
]
private static void registerContextBeans(ConfigurableListableBeanFactory
beanFactory, DubboSpringInitContext context) {
   // register singleton
   registerSingleton(beanFactory, context);
   registerSingleton(beanFactory, 
   context.getApplicationModel());
   registerSingleton(beanFactory, 
   context.getModuleModel());
}
\end{lstlisting}

\fbox{
    \begin{minipage}{8.5cm}
        \textbf{Ground Truth Warning:} object returned by ``context.getApplicationModel()'' could be null and is dereferenced by call to ``registerSingleton(...)'' at line 6.
    \end{minipage}
}

\fbox{
    \begin{minipage}{8.5cm}
        \textbf{{\tool}'s Explanation:} The values of ``context.getApplicationModel()'' and ``context.getModuleModel()'' are being checked for null before being registered as singletons, so there is no null dereference bug in this code.
    \end{minipage}
}

\end{center}

\caption{An example of a missing \textit{Null Dereference} bug by SkipAnalyzer. In this case, the ``registerSingleton'' is a function declared as outside of the method where the \textit{Null Dereference} is detected. Consequently, {\tool} attempts to establish a connection between the registering for singleton and the \textit{Null Dereference} issue, even though they are not directly related.}
\label{fig:discussion-nd3}

\end{figure}

%% file: figure/discussion/ND4.tex
\begin{figure}[t!]
\begin{center}
    
\begin{lstlisting}[
    numberstyle=\bfseries\small, % Make line numbers bold and small
    numbersep=0pt,   % Adjust the separation between code and line numbers
    numbers=left,
    basicstyle=\small, %or \small or \footnotesize etc.
]
private void cacheNormalInvokers(BitList<Invoker<T>> 
   invokers) {
    BitList<Invoker<T>> clonedInvokers = invokers.clone();
    clonedInvokers.removeIf((invoker) ->
    invoker.getUrl().getProtocol().equals(MOCK_PROTOCOL));
    normalInvokers = clonedInvokers;
}
\end{lstlisting}

\fbox{
    \begin{minipage}{8.5cm}
        \textbf{Ground Truth Warning:} object returned by ``invoker.getUrl().getProtocol()'' could be null and is dereferenced at line 5.
    \end{minipage}
}

\fbox{
    \begin{minipage}{8.5cm}
        \textbf{{\tool}'s Explanation:} The `normalInvokers` object is assigned the value of `clonedInvokers`, which is a cloned version of `invokers`. However, there is no null check for `clonedInvokers` before assigning it to `normalInvokers`. If `invokers.clone()` returns null, it will result in a null dereference bug.
    \end{minipage}
}

\end{center}

\caption{An example of a missing \textit{Null Dereference} bug by {\tool}. In this example, {\tool} missed the \textit{Null Dereference} issue in the method call chain and did not report anything about it.}
\label{fig:discussion-nd4}
\end{figure}

%% file: figure/discussion/RL1.tex
\begin{figure}[h]
\begin{center}
    
\begin{lstlisting}[
    numberstyle=\bfseries\small, % Make line numbers bold and small
    numbersep=0pt,   % Adjust the separation between code and line numbers
    numbers=left,
    basicstyle=\small, %or \small or \footnotesize etc.
]
private void executeForSdkArchive() {
    try (TempDirectory tempDir = new 
    TempDirectory(getClass().getSimpleName());
        ZipFile asarZip = new 
        ZipFile(getSdkArchivePath().get().toFile())) {
      Path modulesPath = 
      tempDir.getPath().resolve(EXTRACTED_SDK);
      try (ZipFile modulesZip = getModulesZip(asarZip, 
      modulesPath)) {
        SdkAsarValidator.validateModulesFile(modulesZip);
        SdkAsar sdkAsar = SdkAsar.buildFromZip(asarZip, 
        modulesZip, modulesPath);
        generateAppApks(sdkAsar.getModule(), tempDir);
      }
    } catch (ZipException e) {
      throw CommandExecutionException.builder()
          .withInternalMessage("ASAR is not a valid zip file.")
          .withCause(e)
          .build();
    } catch (IOException e) {
      throw new UncheckedIOException("An error occurred when
      processing the SDK archive.", e);
    }
  }
\end{lstlisting}

\fbox{
    \begin{minipage}{8.5cm}
        \textbf{Ground Truth Warning:} resource of type ``java.util.zip.ZipFile'' acquired by call to ``ZipFile(...)'' at line 4 is not released after line 20.
    \end{minipage}
}

\fbox{
    \begin{minipage}{8.5cm}
        \textbf{{\tool}'s Explanation:} The resources ``tempDir'' and ``asarZip'' are allocated using the try-with-resources syntax, which means they will be automatically released and closed after the try block finishes. Therefore, there is no resource leakage bug in this code.
    \end{minipage}
}

\end{center}

\caption{An example of a missing \textit{Resource Leak} bug by {\tool}. In this example, Although {\tool} correctly identify ``tempDir'' as not buggy, it cannot detect the ``asarZip'', which is an actual \textit{Resource Leak} bug. The code does not use Java 7 try-with-resources (try(...)) syntax for the ``asarZip'' object. It only uses it for ``tempDir'' and also ``modulesZip'' in the next lines.}
\label{fig:discussion-rl1}
\end{figure}

%% file: sec/related.tex
\section{Threats to Validity}
\label{sec:threats}

In this research, our focus is solely on Java projects, and we do not consider projects developed in other programming languages. This choice stems from the limitation that Infer, the tool we employ for generating warnings and the dataset, is not capable of handling multiple programming languages. However, it is worth noting that Infer can be used with languages like C and C\#. Therefore, conducting a similar analysis on additional languages that Infer is compatible with can contribute to the validity of our research.

Another potential challenge to the validity of our work is that the evaluation is conducted exclusively on a dataset created from warnings generated by Infer, without taking into account warnings generated by alternative static analysis tools like SpotBugs\cite{spotbugs-wild} or Error-Prone\cite{error-prone}. In this study, we opted not to include SpotBugs and Error-Prone as static bug detectors. This decision was made because these tools categorize bug types differently, and we aimed to maintain consistency in our bug classification \cite{harzevili2022characterizing}. Additionally, Infer has better precision and outperforms them in accuracy \cite{kharkar2022learning}.

Also, we exclusively employed LLM models from OpenAI, specifically ChatGPT-4 and ChatGPT-3.5 Turbo. It is essential to recognize that other companies have also introduced their LLM models, such as Meta's Llama2 and Google's PaLM2 and Bard. These alternative models may bring their own unique features and performance characteristics, which could potentially impact the validity of our findings.

Furthermore, it is important to acknowledge that the performance of our {\tool} may exhibit variations across different sets of projects. To account for this variability and enhance the generalizability of our results, we have included a diverse array of projects from various repositories and backgrounds.


\section{Related Work}
\label{sec:related}


\subsection{LLM Applications in Software Engineering} 
Recently, there has been a considerable volume of research dedicated to exploring the capabilities of Large Language Models (LLMs) within the domain of Software Engineering (SE). For example, several studies focus on automated program repair \cite{apr-era-llm, codex-apr-improving, apr-llm-multilingual, jin2023inferfix, apr-of-programs-from-llms, llm-apr-splash, vulrepair, apr-clm-llm, chatgpt-promptdesign-dl-repair, alpha-repair, hadi-codebert}. For instance, Xia et al.~\cite{apr-era-llm} conducted the first empirical study to evaluate nine recent state-of-the-art LLMs for automated program repair tasks on five different repair datasets. Mashhadi and Hemmati~\cite{hadi-codebert} fine-tuned a repair dataset of single-line Java bugs on the CodeBERT model. Also, numerous studies in software testing and fuzzing utilize LLMs \cite{llm-fewshot-tester, codemosa-coverage-plateu, fuzzgpt, titan-fuzz}. For example, Deng et al. \cite{fuzzgpt} proposed FuzzGPT, a novel LLM-based fuzzer that can produce unusual programs for fuzzing real-world systems. The same authors also provide TitanFuzz \cite{titan-fuzz}, the first LLM-based approach for fuzzing Deep Learning (DL) libraries. Kang et al. \cite{llm-fewshot-tester} proposed LIBRO, a framework that uses LLMs to automate test generation from general bug reports. Furthermore, there are studies that focus on Oracle generation \cite{oracle1, oracle2, oracle3}. For example, Tufano et al. \cite{oracle1} proposed a novel assertion generation approach using a BART transformer model. Nie et al. \cite{oracle2} proposed an approach for predicting the next statement in test methods that need reasoning about the code execution by utilizing CodeT5 model. Dinella et al. \cite{oracle3} introduced TOGA, a neural transformer-based method for inferring both exceptional and assertion test conditions by considering the context of the main method. Also, there are studies that focus on mobile applications such as Android. \cite{prompting-is-all-you-need, android2}. Feng et al. \cite{prompting-is-all-you-need} introduced an automated technique for Android bug reproduction, utilizing ChatGPT-3.5 to extract reproducible steps and automate the bug replay procedure. Taeb et al. \cite{android2} proposed an automated approach for producing a navigable video from a manual accessibility test of a mobile application  by utilizing ChatGPT-4.

%% file: sec/conclusion.tex
\section{Conclusion}
\label{sec:conclusion}

This paper introduces a novel LLM-based tool for static code analysis. Our developed tool, {\tool}, can showcase the capabilities of LLMs like ChatGPT models in carrying out code analysis tasks, including static bug detection, false-positive warning removal, and bug repair. To generate warnings, we employed Infer, a well-established static analysis tool, on prominent open-source Java projects and projects from prior research. Subsequently, we meticulously labeled each of the generated warnings for two types of bugs: \textit{Null Dereference} and \textit{Resource Leak}, thereby creating a new dataset for our analytical work. We then harnessed the power of different ChatGPT models (i.e., ChatGPT-3.5 Turbo, ChatGPT-4) under different prompting strategies. Our experiments reveal that {\tool}'s components can outperform baseline counterparts, all while offering significant advantages in terms of reduced costs and complexity. 

In the future, our research endeavors will broaden in scope as we aim to explore a wider array of Large Language Models (LLMs), such as Meta's Llama, Google's Bard, and PaLM2, and integrate them into {\tool}. We also intend to delve into a comparative analysis between fine-tuned LLMs and LLMs that are specialized for specific tasks through prompt engineering. Additionally, we find it intriguing to assess {\tool}'s performance on different types of bugs. Furthermore, adding additional components to {\tool} to further examine coding practices and styles would be helpful.
